\documentclass[a4paper,11pt]{article}

\newcommand{\R}{I\!\!R}

\begin{document}

\title{
Finding an ordinary conic and an ordinary hyperplane}

\author{
Olivier Devillers\thanks{INRIA, BP 93, 06902 Sophia Antipolis, France.
  Olivier.Devillers@sophia.inria.fr}
\and
Asish Mukhopadhyay\thanks{Indian Institute of Technology,Kanpur 208016, India.
  am@iitk.ac.in, work of this author 
was done in part while he was visiting INRIA. The support provided
by INRIA, financial and otherwise, is gratefully acknowledged.}
}
\maketitle

\abstract{    

Given a finite set of non-collinear points
in the plane, there exists a line that passes through exactly two
points. Such a line is called an {\em ordinary line}. 
An efficient algorithm for computing such a line was proposed
by Mukhopadhyay et al \cite{mah-olpcg-97}.

In this note we extend this result in two directions. We first show 
how to use this algorithm to compute an 
{\em ordinary conic}, that is, a conic passing through exactly five 
points, assuming that all the points do not lie on the same conic.
Both our proofs of existence and the consequent algorithms are simpler than 
previous ones.
We next show how to compute an ordinary hyperplane in three and higher
dimensions.
}              

\section{Introduction}
 
Let $\cal S$ be a set of $n$ points in the plane.
A {\em connecting} line of $\cal S$ is a line that passes through
at least two of its points. 
A connecting line is said to be {\em ordinary} if it
passes through exactly two points of $\cal S$.  

The problem of establishing the existence of such a line 
originated with Sylvester
\cite{s-mq1-93}, who proposed the following problem in 1893: 

\begin{quote}
If $n$ points in the
plane are such that a line passing through any two of 
them passes through a third point, then are the points collinear? 
\end{quote}


No solution came forth during the next forty years. In 1943,
a positive version of the same problem was proposed by Erdös
\cite{e-pn4-44}, 
and in the following year a solution by Gallai \cite{g-spn4-44} 
appeared in print.

Subsequently other proofs also appeared, notable among which 
were the proofs by Steinberg \cite{s-spn4-44} and 
Kelly \cite{km-noldn-58}.
These results showed that the answer is in the affirmative for 
real projective geometry in the plane. 
Therefore if the points of $\cal S$ are not collinear then there is at 
least one ordinary line.
In fact, Kelly and Moser \cite{km-noldn-58} showed that there are 
at least $3n/7$ ordinary lines.

A set of points is said to be {\em co-conic} if all the points
lie on one conic.
In this paper we address a more general version of the ordinary line
problem: given a set of $n$ points in the plane that are not co-conic, find
a conic that passes through exactly five points.

Our algorithm provides a constructive proof of the existence 
of such a conic. Another proof is contained in \cite{ww-stcs-88}.
Our proof is very simple and allows us to relate a result on the number
of ordinary lines to the number of ordinary conics.

The paper is organized as follows. In the next section we 
discuss some mathematical preliminaries. The algorithm is
discussed in the third section. We conclude in the fourth and
final section.

\section{Preliminaries}

\subsection{Notations and basic results}

\paragraph{Space of conics}
Let $\cal S$ be a set of $n$ points in $\R^2$.
Let $\phi$ be the transformation that maps a point $p=(x,y)\in\R^2$ to 
the point  $p^{\star}=(x^2,y^2,xy,x,y)\in\R^5$.

Under this transformation, ${\R^2}^{\star}$ is the 2 dimensional manifold image
of $\R^2$ and ${\cal S}^{\star}$ the map of $\cal S $in $\R^5$ . 

If $\cal C$ is a conic in $\R^2$ with the equation
$ax^2+by^2+cxy+dx+ey+f=0$, then 
$\phi({\cal C})=\cal C^{\star}$ is the intersection of
$\phi({\R^2})={\R^2}^{\star}$
with the hyperplane ${\cal C}^v$:
$au+bv+cw+dx+ey+f=0$ in  $\R^5$.
We identify the conics of ${\R^2}$ to hyperplanes of
${\R^5}$, which can be called {\em space of conics}.

This idea of mapping points in five dimensions is a natural
generalization of the usual space of circles widely used in
computational geometry \cite{p-gcc-70,e-acg-87,dmt-ssgtu-92i}
that associates a circle $\cal C$
in the plane to a point ${\cal C}^{\bullet}$ in three-dimensions
and to the polar hyperplane ${\cal C}^v$
of ${\cal C}^{\bullet}$ with respect to the unit paraboloid.

\paragraph{Flats}
We recall a few basic results
 from finite-dimensional vector spaces.  

A flat $F$ is an affine subspace of $\R^5$
such that for any two points $p,q \in F$, $\alpha p + \beta q \in F$,
where $\alpha + \beta = 1$.  

A flat is defined by one of its point and its direction
 $\vec{F}=\{p-q, \; p,q\in F\}$ which is
a vectorial subspace of the vectorial space $\R^5$
(a flat is a set of points, its direction a set of vectors).

Two subspaces $\vec{F_1}$ and $\vec{F_2}$ of $\R^5$
are called supplementary if and only if 
$\vec{F_1}\cap \vec{F_2}=\{0\}$ and $dim(S_1)+dim(S_2)=5$.

Two flats $F_1$ and $F_2$ having supplementary directions have an
unique intersection point.

If $A$ and $B$ are two subsets of $\R^5$,
we define the affine hull $A\oplus B$ as the
smallest flat that contains both $A$ and $B$.

\paragraph{Point-hyperplane duality\label{duality}}

Point-hyperplane duality is a common transformation
in computational geometry \cite{e-acg-87,ps-cgi-90}.
A point $p$ at distance $r$ from the origin $O$ is associated with
the hyperplane normal to $Op$ at distance $1/r$ from the origin.

This transformation reduces the problem
of computing the intersection of a finite set of half-spaces, each
containing the origin, to the problem of computing the convex
hull of the corresponding points in dual space.

\paragraph{Inversion\label{inversion}}
An inversive transformation maps
a point $p$ at distance $r$ from the origin, $O$, to the point $p'$
at distance $1/r$ from $O$, lying on the half-line $[Op)$
\cite{ps-cgi-90}.

This involutary
transformation has the interesting properties that the images
of spheres  and hyperplanes are spheres or  hyperplanes.
Particularly, spheres passing through $O$ are exchanged with hyperplanes.

\subsection{Ordinary line\label{oline}}
For completeness, we briefly sketch  the algorithm for finding
an ordinary line in a finite set of non-collinear and coplanar points.

Let $l$ be a directed line (direction $\vec{v}$) through exactly one 
point $p_0$ of $\cal S$. Let $q_{\lambda}=p_0+\lambda \vec{v}$.
We find the line passing through at least two
points of $\cal S$ that cuts $l$ in
a point $q_{\lambda}$ with minimal $\lambda >0$.
Such a line passes through two points consecutive in polar order around
$p_0$ and can thus be found in $O(n\log n)$ time.
Either this line is ordinary or a line through $p_0$
and a point on this line is ordinary. For details see 
Mukhopadhyay et al. \cite{mah-olpcg-97}.     

\section{Algorithm}

The idea behind the algorithm is to find a hyperplane that 
passes through exactly five points of $\cal S^{\star}$.
In the $\R^2$ plane this corresponds to a conic that 
passes through exactly five points of $\cal S$.

We first find a conic that passes through exactly three points
of $\cal S$. We do this as follows.
We choose $p,q,r\in \cal S$ and $s,t \not\in \cal S$ such that
no three (four) of the five points are collinear.
Denote by $\vec{\imath}$ the vector $(1,0)\in\R^2$ and consider
the conic ${\cal A}_{\theta}$ passing through the five points
$p,q,r,s,t+\theta\vec{\imath}$.

For any point $\rho\in\cal S$, there exist at most two values
$\theta =\theta_{\rho}$ or $\theta =\theta_{\rho}'$ such that 
$\rho\in{\cal A}_{\theta}$. This is because if

\[a_{\theta}x^2 + b_{\theta}y^2 + c_{\theta}xy + d_{\theta}x +
e_{\theta}y + f_{\theta} = 0\]

\noindent
is the conic that passes through the points, 
$p,q,r,s,t+\theta\vec{\imath}$ then each of the coefficients 
is of second degree in $\theta$.
So it is easy to determine some
$\theta_0$ different from all these values such that 
${\cal A}_{\theta_0} \cap {\cal S}=\{p,q,r\}$.

Now the affine hull ${\cal B}=p^{\star} \oplus q^{\star} \oplus r^{\star}$
is a subset of the hyperplane 
${\cal A}_{\theta_0}^{\star}$ and so is the affine hull spanned by 
the points $p^{\star}$, $s^{\star}$ and 
$(t+\theta_0\vec{\imath})^{\star}$.

\begin{figure}[t] 
\begin{center}
\def\IPEfile{Mapping.ipe}\input{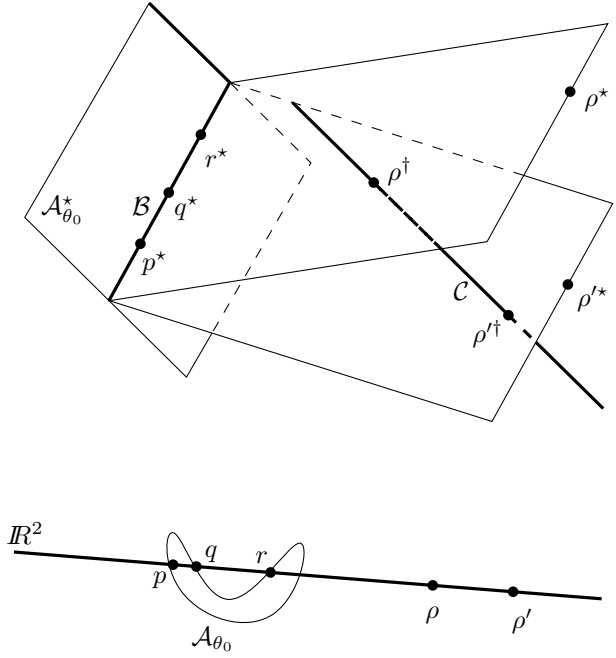} \caption{\label{Mapping}
Mapping in 5 dimensions}
\end{center} \end{figure}

Moreover, these two sets intersect in the single point $p^{\star}$
(see Figure \ref{Mapping}).

Let $\gamma$ be a point of $\R^5$ not in ${\cal A}_{\theta_0}^{\star}$.
Translate the affine hull of the points, $p^{\star}$, $s^{\star}$ and 
$(t+\theta\vec{\imath})^{\star}$ to pass through $\gamma$. 
Let ${\cal C}$ denote this translated affine hull.   

For any point $\rho\in{\cal S} - \{p,q,r\}$, we construct the point
$\rho^{\dag}=({\cal B}\oplus \rho^{\star}) \cap {\cal C}$.
The intersection is exactly one point because the directions of
the flats 
${\cal B}\oplus \rho^{\star}$ and ${\cal C}$ are supplementary
subspaces of the vectorial space $\R^5$. Otherwise, $\rho^{\star}$ 
would belong  to
${\cal A}_{\theta_0}^{\star}$ which is impossible since 
$\rho\not\in{\cal A}_{\theta_0}$ by the definition of $\theta_0$.

The set of points ${\cal S}^{\dag}=\left\{\rho^{\dag};
\rho\in{\cal S}\ - \{p,q,r\}\right\}$
lies in the two-dimensional plane $\cal C$.
Let $l$ be a line in that plane and $\cal H$ the hyperplane through 
$\cal B$ and $l$.
By construction $\rho^{\dag}\in {\cal H}$ if and only if 
$\rho^{\star}\in {\cal H}$;
indeed by the definition of $\rho^{\dag}$,
the line $(\rho^{\dag}\rho^{\star})$ cuts $\cal B$ (in $\rho^{\ddag}$)
and thus an hyperplane ${\cal H}$ cannot contains
$\cal B$ ($\Rightarrow \rho^{\ddag}\in{\cal H}$)
and $\rho^{\dag}$ without containing $\rho^{\star}$.
 Thus at this point we see
a complete equivalence between the problem of finding an ordinary line
for ${\cal S}^{\dag}$ and an ordinary conic for $\cal S$ through 
the points $p$, $q$ and $r$.

If all the points of ${\cal S}^{\dag}$ are collinear, then all 
the points of ${\cal S}^{\star}$
are in the same hyperplane and hence all the points of $\cal S$ are 
co-conic.

Otherwise, there exists an ordinary line $l$ in $\cal C$, 
and the corresponding hyperplane $\cal H$
contains only five points of  ${\cal S}^{\star}$. The corresponding 
conic is ordinary and passes through $p,q,r$.

The following theorem is a consequence of the above discussion and
the fact that there are at least $\frac {3n}{7}$ ordinary lines:

\begin{quote}
{\bf Theorem:} \em
Given a set $\cal S$ of $n$ points in the plane that
 are not co-conic,
then for any three non collinear points in $\cal S$,
there exist at least $\frac{3(n-3)}{7}$ ordinary conics
of $\cal S$ that pass though these three points.
Furthermore, such an ordinary conic can be found in $O(n\log n)$ time.
\end{quote} 

\section{Ordinary plane in three dimensions}

Given a set ${\cal S}$
of $n(\geq 3)$ points in three space, a connecting plane (that is, a plane
through some three points of $\cal S$) is defined
to be ordinary if all but one of the points of
${\cal S}$ that lie on it are collinear.
Such a plane always exists, unless all the points of 
${\cal S}$ are collinear.

A plane that passes through exactly three points is certainly 
ordinary in the sense of this definition;
however, such a 
plane need not exist.

As an example, place three or more points on each of two skew lines
in three space. This configuration of points has no connecting plane
that is defined by exactly three points, all the
ordinary planes contain one of the two lines,
(see Motzkin \cite{m-lpcpf-51}).

We show that the ideas sketched in Section~\ref{oline} can be 
generalized to three and higher dimensions to compute a plane that is 
ordinary in the sense of the above definition. 

Let $p_0$ be a point of $\cal S$ and $\gamma$ a line through $p_0$.
Let $\Lambda$ be a plane through three points $p_1,p_2,p_3$ of
$\cal S$ such that its distance to $p_0$, measured along $\gamma$, is 
minimum among all possible connecting planes of $\cal S$ that intersect
$\gamma$.

If $\Lambda$ is ordinary, we are done; otherwise, set
$g=\gamma\cap\Lambda$ and let $\Gamma$ be an arbitrary plane 
containing $\gamma$; finally, set $\lambda =\Lambda\cap\Gamma$.

Let $p_1$, $p_2$, $p_3$ and $p_4$ be points of $\cal S$ in $\Lambda$
such that no 3-tuple of the form $p_1p_ip_j$, $i,j \in \{2,3,4\}$, are 
collinear (such points exist in $\Lambda$ since $\Lambda$ is not 
ordinary).
We consider the planes through $p_0p_1$ and $p_2$,
$p_3$, $p_4$ respectively. Let $L_2$, $L_3$ and $L_4$ be their 
respective intersections with
$\Gamma$ and $l_2$, $l_3$ and $l_4$ their respective intersections 
with $\lambda$.

\begin{figure}[t] \begin{center}
\def\IPEfile{ThreeD.ipe}\input{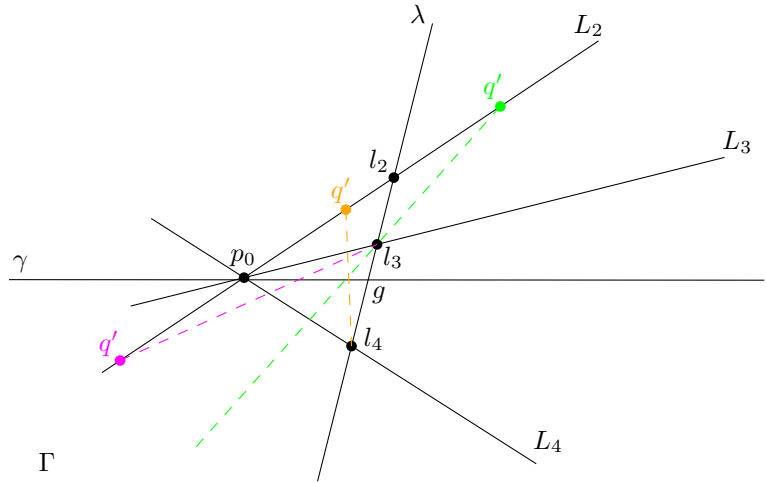} \caption{\label{ThreeD}
Ordinary plane in three dimensions}
\end{center} \end{figure}

Assume, without loss of generality, that $l_2$ is separated from $g$
along $\lambda$ by $l_3$ or $l_4$.
Then we claim that the plane determined by the points $p_0, p_1, p_2$
is ordinary. That is all the points, barring $p_0$, are on the line
determined by $p_1, p_2$. 

If not, let $q$ be a point, distinct from 
$p_0$, lying outside this line.
In fact, if $ q \not\in \Lambda\cap L_2$, one of the two planes 
$qp_1p_3$ or $qp_1p_4$ must pass between $g$ and $p_0$, contradicting 
the definition of $\Lambda$.
 The dashed lines
on  Figure \ref{ThreeD} show the different cases for the intersection
of the plane with $\Gamma$ depending of the position of
$q'=(p_1q)\cap L_2$.

It remains to find $\Lambda$ efficiently. It is clear that
if we consider the cell in the arrangement of
the $O(n^3)$ planes defined by points of ${\cal S}- \{p_0\}$,
that contains $p_0$, $\Lambda$ is incident on the facet that is hit by 
$\gamma$.

By a point-plane duality transformation, with $p_0$ as  center,
the cell containing $p_0$
is mapped into the convex hull of the $O(n^3)$ vertices of an 
arrangement of $n-1$ planes (see paragraph \ref{duality}).


In two dimensions we have to compute the convex hull of the vertices of
an arrangement of $n-1$ lines and it is not difficult to see that a 
vertex can be on the convex hull only if the two lines have 
consecutive slopes (in the set of all slopes).
In three dimensions, the phenomenon is similar if we consider the 
Gaussian diagram of the normals to the $n-1$ planes
(the convex hull of the unit normal vectors to $n-1$ planes)
\cite{ab-scsp-87,gs-ccrs-87,l-spcsa-83}.
Three planes define
a vertex of the convex hull only if their normal vectors define a face
of the Gaussian diagram.
Since the Gaussian diagram can be computed in $O(n\log n)$ time,
we get the following result:

\begin{quote}
{\bf Theorem:} \em
If $\cal S$ a set of $n$ non-coplanar points in 3-space,
then an ordinary plane can be found in $O(n\log n)$ time.
\end{quote}

The same ideas extend to higher dimensions. We can find an ordinary 
hyperplane with the help of a Gaussian diagram. The complexity is 
identical to the complexity of computing the convex hull in that 
dimension.

\section{Conclusions}

In this note we have shown that an algorithm for finding
a line through exactly two points of a given set of non-collinear
points can be used to find a conic through exactly 
five points, if we assume that all the points are not co-conic.

It is particularly easy to find an ordinary circle passing through a 
chosen point, $p$, of the given set of points, if we allow for a 
degenerate circle. We simply apply an inversion transformation 
with $p$ as the center of inversion (see paragraph \ref{inversion}).
Solve the ordinary line problem for the remaining $n - 1$ 
transformed points. We have a degenerate circle if the ordinary 
line found passes through $p$, else its image is an
ordinary circle passing through $p$. We also conclude that 
at least $3(n-1)/7$ ordinary circles pass through a chosen point.

By applying a stereographic projection we note that if $n$ points
on a real sphere do not lie in the same plane then there is plane
containing exactly three of them.

\end{document}